\documentstyle[11pt,newpasp,twoside,graphicx,psfig]{article}
\def\edcomment#1{\iffalse\marginpar{\raggedright\sl#1\/}\else\relax\fi}
\reversemarginpar

\begin{document}
\title{Young star clusters: Progenitors of globular clusters!?}
 \author{Peter Anders, Uta Fritze - v. Alvensleben}
\affil{Universit\"ats-Sternwarte G\"ottingen, Geismarlandstrasse 11, 37083 G\"ottingen, Germany}
\author{Richard de Grijs}
\affil{Department of Physics \& Astronomy, The University of Sheffield, Hicks Building, Hounsfield Road, Sheffield S3 7RH, UK}

\begin{abstract}
Star cluster formation is a major mode of star formation in the extreme conditions of interacting galaxies and violent starbursts. Young clusters are observed to form in a variety of such galaxies, a substantial number resembling the progenitors of globular clusters in mass and size, but with significantly enhanced metallicity. From studies of the metal-poor and metal-rich star cluster populations of galaxies, we can therefore learn about the violent star formation history of these galaxies, and eventually about galaxy formation and evolution.
We present a new set of evolutionary synthesis models of our GALEV code, with special emphasis on the gaseous emission of presently forming star clusters, and a new tool to compare extensive model grids with multi-color broad-band observations to determine individual cluster masses, metallicities, ages and extinction values independently. First results for young star clusters in the dwarf starburst galaxy NGC 1569 are presented. The mass distributions determined for the young clusters give valuable input to dynamical star cluster system evolution models, regarding survival and destruction of clusters. We plan to investigate an age sequence of galaxy mergers to see dynamical destruction effects in process.
\end{abstract}

\section{Models \& tests}
We use our evolutionary synthesis code to study the basic physical parameters of star clusters in interacting galaxies and violent starbursts. The main ingredients are: stellar isochrones from the Padova group for metallicities in the range -1.7 $\le$ [Fe/H] $\le$ +0.4, a stellar initial mass function (usually assumed to be Salpeter-like), the library of stellar spectra by Lejeune et al. (1997, 1998), and gaseous emission (both lines and continuum). From the integrated spectrum we derive magnitudes in a large number of filter systems. The gaseous emission contributes significantly to the integrated light of stellar populations younger than $3 \times 10^7$ yr both in terms of absolute magnitudes and derived colors (see Fig. 1 and Anders \& Fritze - v. Alvensleben 2003). The updated models are available from \begin{center}
{\bf http://www.uni-sw.gwdg.de/$\sim$galev/panders/} \end{center}

While star clusters can easily be approximated as simple stellar populations (SSPs; with all stars having the same age, metallicity and extinction), more complicated star formation / metallicity evolution histories of galaxies can be studied by superimposing appropriate SSPs.\\

\begin{figure}
\begin{center}
\includegraphics[angle=-90,width=0.45\linewidth]{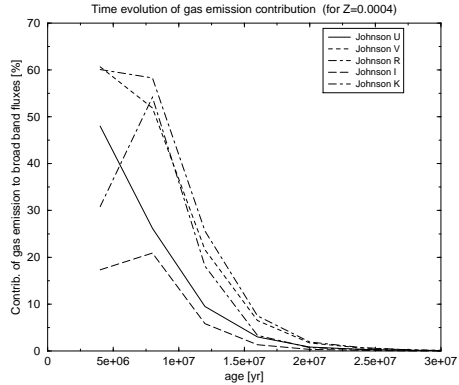}
\end{center}
\caption{Contribution of gaseous emission to various standard Johnson magnitudes for metallicity Z = 0.0004 as a function of time.}
\end{figure}

\begin{figure}
\begin{center}
\includegraphics[angle=-90,width=0.62\linewidth]{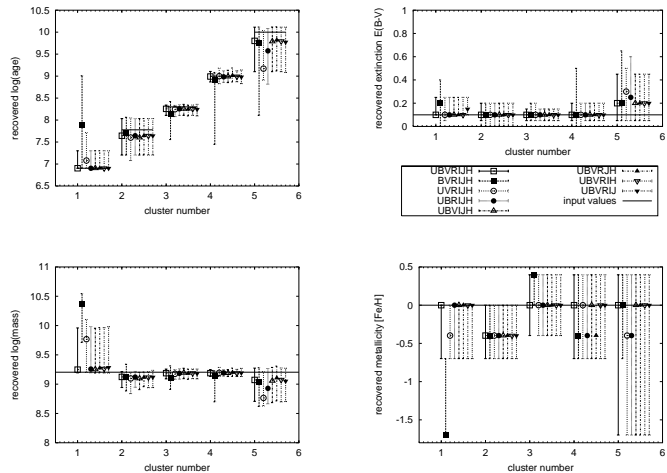}
\end{center}
\caption{Dispersion of recovered properties of artificial clusters, assuming availability of {\sl UBVRIJH} and using passband combinations consisting of 6 out of the available 7 passbands for the analysis, as indicated in the legend. Input parameters for the artificial clusters are: solar metallicity, E(B-V) = 0.1 mag, ages = 8, 60, 200 Myr, 1, 10 Gyr.}
\end{figure}

We developed a tool to compare our evolutionary synthesis models with observed cluster SEDs to determine the cluster parameters age, metallicity, internal extinction and mass independently. Using artificial clusters with various input parameters (with SEDs taken directly from the evolutionary synthesis models) we systematically studied the impact of the choice of passbands, of finite observational photometric uncertainties, and {\sl a priori} assumptions on our analysis results. One example of these tests is shown in Fig. 2. Additional tests were performed using broad-band observations for star clusters in NGC 3310 (de Grijs et al. 2003a,b), confirming the results from the artificial cluster tests. Due to the young age of this cluster system these additional tests are restricted to ages younger than approx. 200 Myr. From our tests we conclude that:
\begin{enumerate}
\item At least 4 passbands are necessary to determine the 3 free parameters age, metallicity and extinction, and the mass by scaling the SED, independently.
\item The most important passbands are the {\sl U} and {\sl B} bands; for systems older than roughly 1 Gyr the {\sl V} band is equally important.
\item NIR bands significantly improve the results by constraining the metallicity efficiently.
\item A wavelength coverage as long as possible is desirable. Best is UV through NIR, thus tracing the pronounced kink/hook in the SEDs around the {\sl B} band.
\item Large observational errors and/or wrong {\sl a priori} assumptions may lead to completely wrong results.
\end{enumerate}

\vspace{-0.4cm}
\section{The case of NGC 1569}
As a first application the dwarf starburst galaxy NGC 1569 was chosen. Our sample enlarges the number of observed star clusters in this galaxy by a factor of 3. Our results for the physical parameters are in agreement with previous results of the starburst history in NGC 1569 in general, and of the two prominent ``super star clusters'' in particular, regarding age, mass and metallicity. In addition, we find a surprising change in the cluster mass function with age: The clusters formed during the onset of the burst (approx. 25 Myr ago) seem to exhibit an excess of massive clusters as compared to clusters formed more recently. Using various statistical methods we show the robustness of this result (Anders et al. 2003b).

\vspace{-0.4cm}
\section{References}
Anders, P., Fritze - v. Alvensleben, U. 2003, \aap, 401, 1063\\
Anders, P., Bissantz, N., Fritze - v. Alvensleben, U., de Grijs, R. 2003a,\\ \mnras, {\sl submitted}\\
Anders, P., de Grijs, R., Fritze - v. Alvensleben, U., Bissantz, N. 2003b,\\ \mnras, {\sl submitted}\\
de Grijs, R., Fritze - v. Alvensleben, U., Anders, P., Gallagher, J. S., Bastian, N., Taylor, V. A., Windhorst, R. A. 2003a, \mnras, 342, 259\\
de Grijs, R., Anders, P., Bastian, N., Lynds, R., Lamers, H.J.G.L.M., O'Neil, Jr., E.J. 2003b, \mnras, 343, 1285\\
Lejeune, T., Cuisinier, F., Buser, R. 1997, \aaps, 125, 229\\
Lejeune, T., Cuisinier, F., Buser, R. 1998, \aaps, 130, 65
\end{document}